\documentclass[twocolumn,prb]{revtex4}
\usepackage{amsfonts}
\usepackage[T1]{fontenc}
\usepackage{amsmath,amsbsy,amssymb,graphicx}
\usepackage{times}

\begin{document}

\title{Chiral edge soliton in nonlinear Chern systems}
\author{Motohiko Ezawa}
\affiliation{Department of Applied Physics, University of Tokyo, Hongo 7-3-1, 113-8656,
Japan}

\begin{abstract}
We study the effect on the chiral edge states by including a nonlinearity to
a Chern insulator which has two chiral edge states with opposite
chiralities. We explore a quench dynamics by giving a pulse to one site on
an edge and analyzing the time evolution of a wave packet. Without the
nonlinearity, an initial pulse spreads symmetrically and diffuses. On the
other hand, with the nonlinearity present, a solitary wave is formed by the
self-trapping effect of the nonlinear term and undergoes a unidirectional
propagation along the edge, which we identify as a chiral edge soliton. A
further increase of the nonlinearity induces a self-trapping transition,
where the chiral wave packet stops its motion. It is intriguing that the
nonlinearity is controlled only by changing the initial condition without
changing a sample.
\end{abstract}

\maketitle

\textbf{Introduction}: Topological physics is one of the most active fields
in condensed matter physics, which \ has been applied to photonic \cite%
{Raghu,Hafe2,Hafezi,KhaniPhoto,WuHu,TopoPhoto,Ozawa16,Ley,KhaniSh,Zhou,Jean,Ota18,Ozawa,Ota19,OzawaR,Hassan,Ota,Li,Yoshimi,Kim,Iwamoto21}%
, acoustic\cite{Prodan,TopoAco,Berto,Xiao,He,Abba,Xue,Ni,Wei,Xue2},
mechanical\cite{Lubensky,Chen,Nash,Paul,Sus} and electric circuit\cite%
{TECNature,ComPhys,Hel,Lu,YLi,EzawaTEC,EzawaLCR,EzawaSkin} systems. The
Su-Schrieffer-Heeger (SSH) model\cite{Malko,XiaoSSH,Jean,Parto} and the
Chern insulator\cite{Raghu,Hafe2,Hafezi} are realized in these systems.
Nonlinear topological physics is an emerging field, which is also studied in
photonic\cite{Ley,Zhou,MacZ,Smi,Tulo,Kruk,NLPhoto,Kirch,TopoLaser},
mechanical\cite{Snee,PWLo,MechaRot,Sin}, electric circuit\cite%
{Hadad,Sone,TopoToda} and resonator\cite{Zange} systems. The nonlinear term
is typically introduced by the Kerr effect in photonics\cite{Chris,Szameit}.
The system is described by the nonlinear Schr\"{o}dinger equation. The
simplest model is the nonlinear SSH model\cite%
{Zhou,Gor,Hadad,Tulo,Dob,NLPhoto}.

A soliton is a stable wave packet in nonlinear systems. Edge solitons are
fascinating objects\cite{Ley,Muk,ZhangSoliton} propagating along an edge of
a sample. Solitons are often described by exact solutions in continuum
theory. However, there seems to be no exact solutions in lattice systems due
to the lack of the continuous translational symmetry in general.

The study of quench dynamics is a powerful method to reveal the essence of
nonlinear topological systems\cite%
{TopoToda,MechaRot,NLPhoto,TopoLaser,NLSkin}, where a pulse is given to one
site on an edge\ as the initial condition and its time evolution is
analyzed. It has been employed to reveal topological edge states in one
dimension\cite{TopoToda,MechaRot,NLPhoto,TopoLaser,NLSkin} and topological
corner states in two dimensions\cite{NLPhoto,TopoLaser}. The initial pulse
remains as it is in the topological phase, while it spreads into the bulk in
the trivial phase. However, there is so far no application of this method to
chiral edge states in two dimensions.

In this paper, we investigate the nonlinear effect on the chiral edge states
in a nonlinear Schr\"{o}dinger equation. We study a topological model
describing a Chern insulator, which is realized in coupled resonator optical
waveguides in the case of photonics. We employ the quench dynamics. First,
without the nonlinear term, the initial pulse spreads symmetrically between
the right and left directions. This is because two chiral edge states with
opposite chiralities are present. Next, with the nonlinear term included, a
solitary wave is formed by the self-trapping effect of the nonlinear term
and propagates rightward along the edge. Namely, the chirality emerges due
to the nonlinear term. We may identify it as a chiral edge soliton. In
addition, a self-trapping transition is induced by the self-trapping effect
beyond a certain magnitude of the nonlinearity, where the chiral edge state
stops its motion. 

\textbf{Model}: We investigate a nonlinear topological system described by
the nonlinear Schr\"{o}dinger equation on the square lattice\cite%
{Eil,Kev,Szameit,Chris,Bandres,Dob},%
\begin{equation}
i\frac{d\psi _{n}}{dt}+\kappa \sum_{m}M_{nm}\psi _{m}+\xi \left\vert \psi
_{n}\right\vert ^{2}\psi _{n}=0,  \label{BasicEq}
\end{equation}%
where $n=\left( n_{x},n_{y}\right) $, $M_{nm}$ is a hopping matrix%
\begin{align}
M_{nm}& =e^{i\alpha n_{y}}\left\vert n_{x}+1,n_{y}\right\rangle \left\langle
n_{x},n_{y}\right\vert  \notag \\
& +e^{-i\alpha n_{y}}\left\vert n_{x},n_{y}\right\rangle \left\langle
n_{x}+1,n_{y}\right\vert ,  \notag \\
& +\left\vert n_{x},n_{y}+1\right\rangle \left\langle n_{x},n_{y}\right\vert
+\left\vert n_{x},n_{y}\right\rangle \left\langle n_{x},n_{y}+1\right\vert ,
\label{HopMat}
\end{align}%
and $\kappa $ is the coupling strength.

This system is topological because the hopping matrix (\ref{HopMat}) is well
known to describe the Chern insulator for $\alpha \neq 0,\pi $. It is also
known to describe the quantum Hall effect with $\alpha $ representing the
penetrated flux into a plaquette of the square lattice. This model is also
realized in photonic systems by making coupled resonator optical waveguides%
\cite{Hafe2,Hafezi,Harari,Bandres}, where $\alpha $ represents a gauge flux
in the Landau gauge, and the nonlinear term is introduced by the Kerr effect%
\cite{Szameit,Chris} with intensity $\xi >0$.

We take $\alpha =\pi /2$ explicitly in what follows. In this case, we have\
a four-band model. It is given by 
\begin{equation}
M\left( k_{x},k_{y}\right) =\left( 
\begin{array}{cccc}
2\cos k_{x} & 1 & 0 & e^{-ik_{y}} \\ 
1 & -2\sin k_{x} & 1 & 0 \\ 
0 & 1 & -2\cos k_{x} & 1 \\ 
e^{ik_{y}} & 0 & 1 & 2\sin k_{x}%
\end{array}%
\right) ,  \label{matrixM}
\end{equation}%
in the momentum space.

\begin{figure}[t]
\centerline{\includegraphics[width=0.48\textwidth]{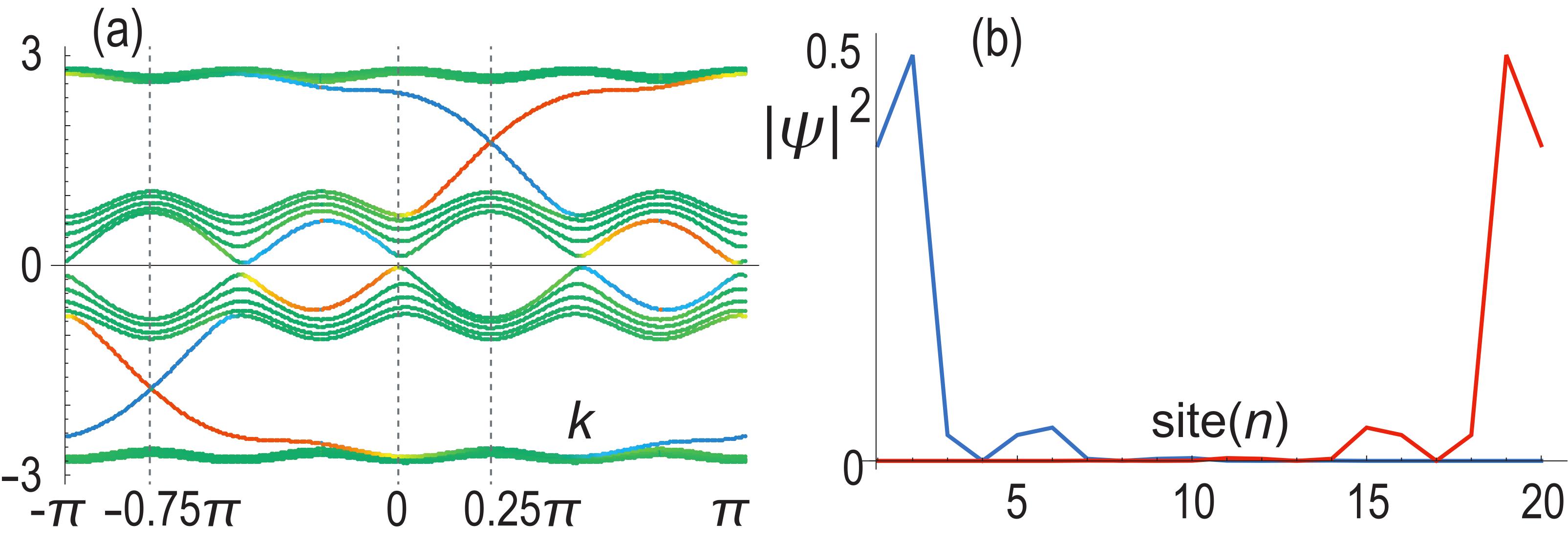}}
\caption{(a) Band structure of the matrix $\protect\kappa M_{nm}$ in
nanoribbon geometry. Curves in red (blue) indicate the left(right)-localized
edge states, while those in green indicate the bulk states. The horizontal
axis is the momentum $k$ ranging from $-\protect\pi $ to $\protect\pi $. The
vertical axis is the energy in units of $\protect\kappa $. (b) The\ LDOS $|%
\protect\psi _{n}|^{2}$ at $k=0.25\protect\pi $. It is found to be localized
at the edges. The horizontal axis is the lattice site $n$. We have taken the
width of the nanoribbon $L=20$.}
\label{FigRibbon}
\end{figure}

As a characteristic feature of a topological system, the topological edge
states emerge in nanoribbon geometry. The band structure of the matrix $%
\kappa M$\ is shown in Fig.\ref{FigRibbon}(a), where we clearly observe four
topological edge states designated by two sets of crossed red and cyan
curves. They are two chiral edge modes with positive energy at $k=0.25\pi $
and negative energy at $k=-0.75\pi $, where the direction of the chirality
is opposite. We show the local density of states (LDOS) at $k=0.25\pi $ in
Fig.\ref{FigRibbon}(b), where there are two edge states localized at the
right and left edges.

\begin{figure}[t]
\centerline{\includegraphics[width=0.48\textwidth]{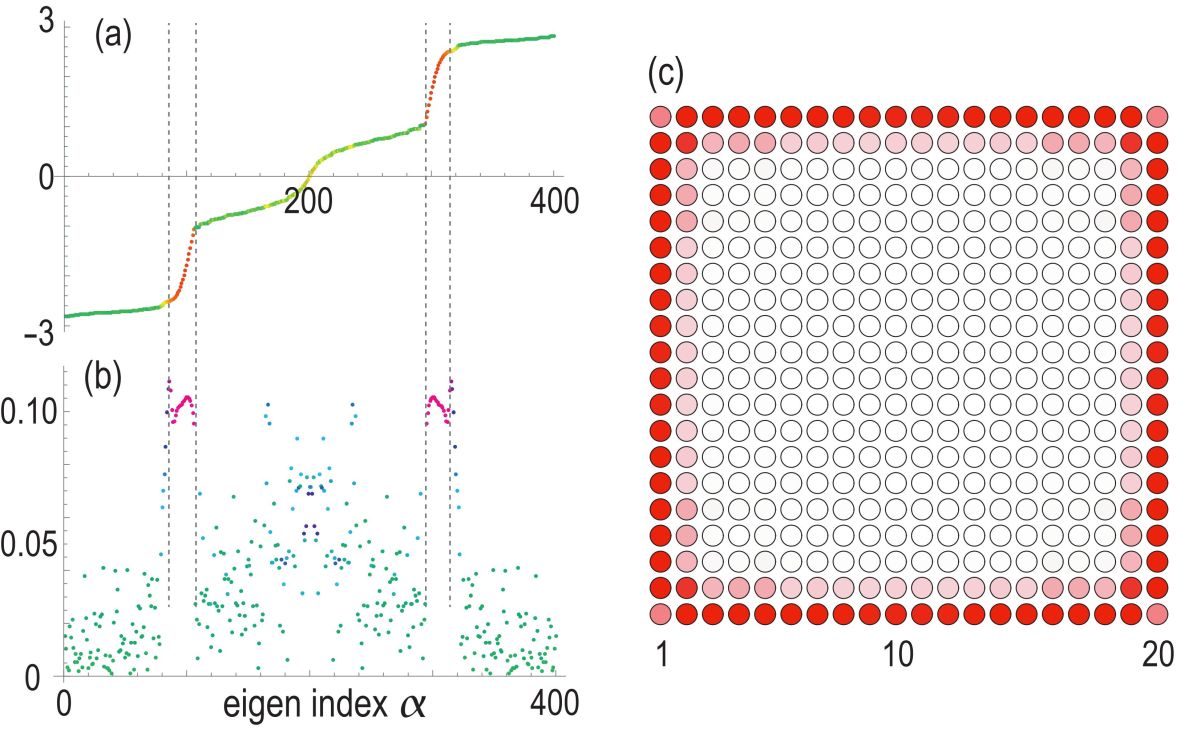}}
\caption{(a) Eigenspectrum of the matrix $\protect\kappa M_{nm}$ in square
geometry. The red parts of a curve indicate the edge states, while the green
parts indicate the bulk states. The horizontal axis is the eigen index $%
\protect\alpha $ of the state $\bar{\protect\psi}_{n}^{\left( \protect\alpha %
\right) }$. The vertical axis is the energy in units of $\protect\kappa $.
(b) The component $|c_{\protect\alpha }|^{2}$ corresponding to the
eigenenergy in (a). (c) The\ LDOS $|\protect\psi _{n}|^{2}$ designated by
the strength of red. It is found to be localized along the edges. We have
used a square with size $20\times 20$. }
\label{FigSquare}
\end{figure}

Next, we calculate the eigenspectrum of the matrix $\kappa M$ in square
geometry, which is shown in Fig.\ref{FigSquare}(a). We also show the LDOS
for an edge state in Fig.\ref{FigSquare}(c), where the eigenfunction is well
localized at the edge of the square.

\textbf{Quench dynamics}: We study a quench dynamics by giving a pulse to
one site on the edge initially and by examining its time evolution. Namely,
we solve the nonlinear Schr\"{o}dinger equation (\ref{BasicEq}) under the
initial condition,%
\begin{equation}
\psi _{0}\left( n_{x},n_{y}\right) =\delta \left(
n_{x}-L_{x}/2,n_{y}-1\right) ,  \label{IniCon}
\end{equation}%
where $L_{x}$\ is the length of the edge along the $x$ axis, $L_{x}$ being
an even number.

\textit{Scale transformation}: By making a scale transformation%
\begin{equation}
\psi _{j}=1/\sqrt{\xi }\psi _{j}^{\prime },
\end{equation}%
it follows from (\ref{BasicEq}) that 
\begin{equation}
i\frac{d\psi _{n}^{\prime }}{dt}+\kappa \sum_{m}M_{nm}\psi _{m}^{\prime
}+\left\vert \psi _{n}^{\prime }\right\vert ^{2}\psi _{n}^{\prime }=0,
\label{DST2}
\end{equation}%
where the nonlinearity parameter $\xi $ is removed.

\begin{figure}[t]
\centerline{\includegraphics[width=0.48\textwidth]{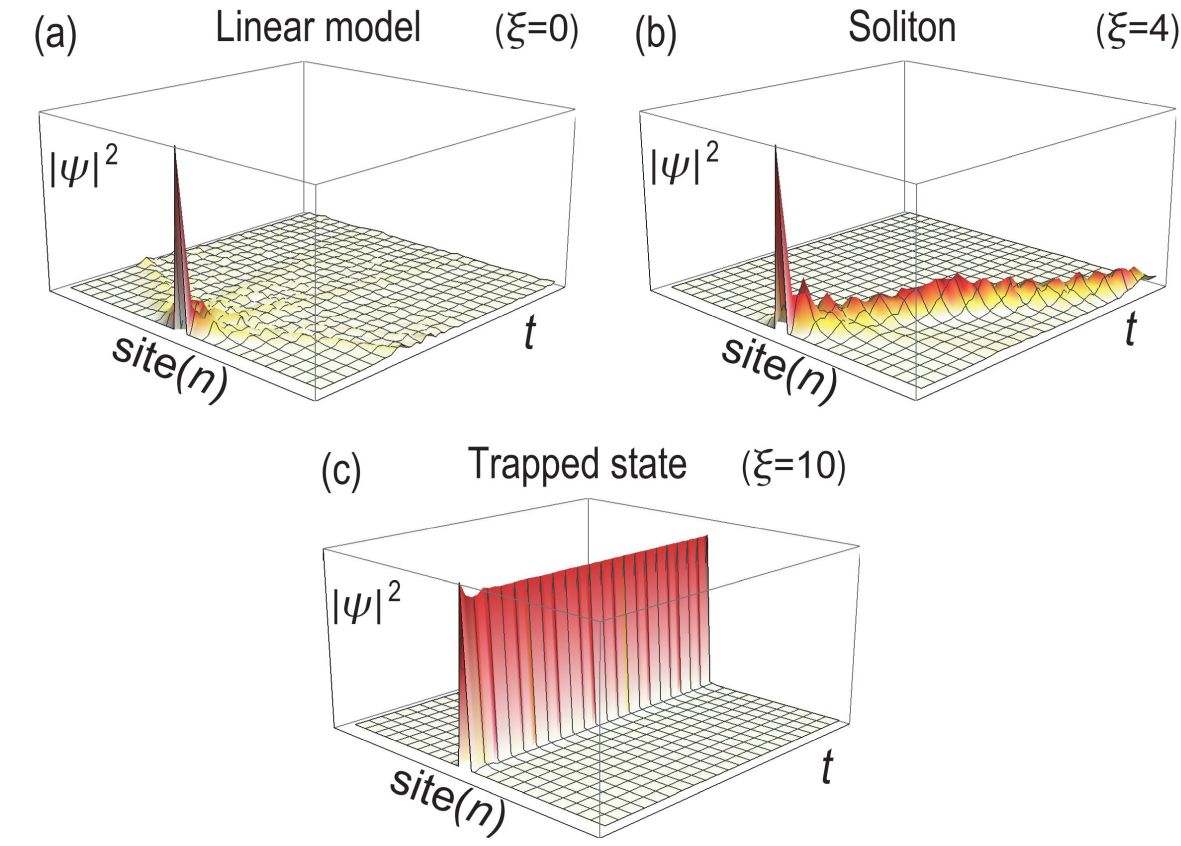}}
\caption{Time evolution of the LDOS $|\protect\psi_n |^{2}$ (a) in the
linear model with $\protect\xi =0$, (b) in a nonlinear model with $\protect%
\xi =4$ and (c) in a strong nonlinear model with $\protect\xi =10$. We have
used a square sample with size $20\times 20$.}
\label{FigPropagate}
\end{figure}

\begin{figure}[t]
\centerline{\includegraphics[width=0.48\textwidth]{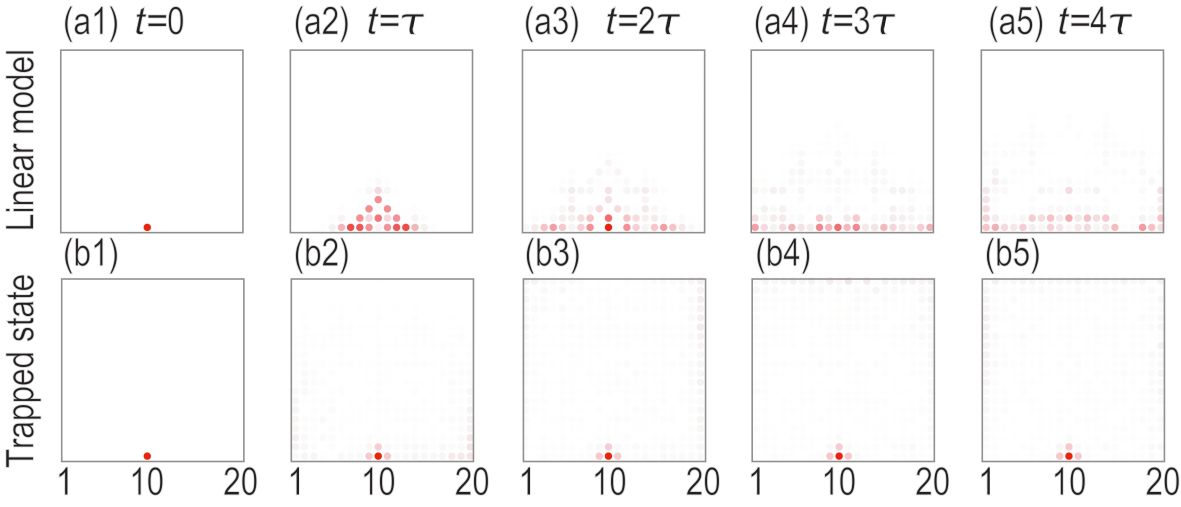}}
\caption{Time evolution of the spatial profile of the LDOS in a whole square
sample with size $20\times 20$. (a1)$\sim $(a8) The LDOS in the linear model
with $\protect\xi =0$, where it diffuses as time passes. (b1)$\sim $(b8) The
LDOS in a strong nonlinear model with $\protect\xi =10$, where it is
unchanged as time passes: The time step is $\protect\tau =2$ in units of $1/%
\protect\kappa $.}
\label{FigQuench}
\end{figure}

We are interested in the dynamics starting from a localized state at $\left(
L_{x}/2,1\right) $ under the initial condition (\ref{IniCon}). This initial
condition is transformed to%
\begin{equation}
\psi _{0}^{\prime }\left( n_{x},n_{y}\right) =\sqrt{\xi }\delta \left(
n_{x}-L_{x}/2,n_{y}-1\right) .  \label{IniCon2}
\end{equation}%
Namely, the quench dynamics subject to Eq.(\ref{BasicEq}) is reproduced with
the use of the nonlinear equation (\ref{DST2}) with the modified initial
condition (\ref{IniCon2}). Consequently, it is possible to use a single
sample to investigate the quench dynamics at various nonlinearity only by
changing the initial condition as in (\ref{IniCon2}).

\begin{figure*}[t]
\centerline{\includegraphics[width=0.98\textwidth]{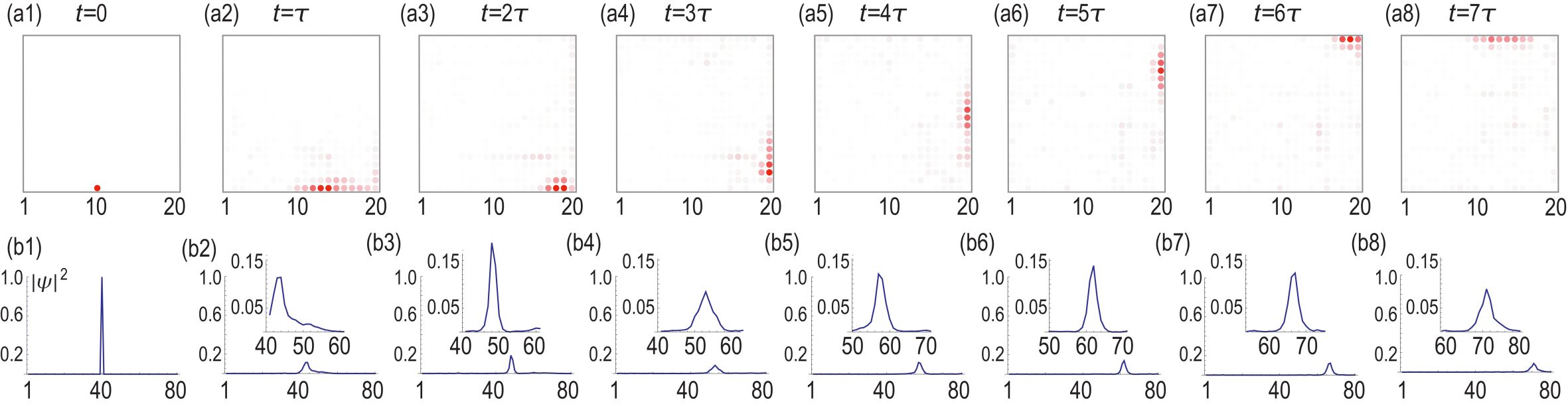}}
\caption{Time evolution of the LDOS $|\protect\psi_n |^{2}$ in the nonlinear
model with $\protect\xi =4$. (a1)$\sim $(a8) Spatial profile of the LDOS in
a whole square sample with size $20\times 20$. (b1)$\sim $(b8) The LDOS
along the edge of a rectangular sample with size $80\times 10$. The time
step is $\protect\tau =10$ in units of $1/\protect\kappa $. Each inset shows
an enlarged figure of a soliton.}
\label{FigChiral}
\end{figure*}

First, we study the linear model by setting $\xi =0$ in Eq.(\ref{BasicEq}).
We expand the initial state (\ref{IniCon}) by the eigenfunctions 
\begin{equation}
\psi _{0}=\sum_{\alpha }c_{\alpha }\bar{\psi}_{n}^{\left( \alpha \right) },
\end{equation}%
where $\bar{\psi}_{n}^{\left( \alpha \right) }$ is the eigenfunction of the
matrix $\kappa M_{nm}$ and $\alpha $ is the index of the eigenenergy,%
\begin{equation}
\kappa M_{nm}\bar{\psi}_{m}^{\left( \alpha \right) }=E_{\alpha }\bar{\psi}%
_{n}^{\left( \alpha \right) }.
\end{equation}%
The square of the component $\left\vert c_{\alpha }\right\vert ^{2}$ is
shown in Fig.\ref{FigSquare}(b). It has peaks at the edge states colored in
red, but it has also values in the bulk states colored in green.

We show the time evolution of the amplitude $\left\vert \psi _{n}\right\vert
^{2}$\ along the edge in Fig.\ref{FigPropagate}, when a pulse is given to
the site $\left( L_{x}/2,1\right) $\ as an initial condition. It exhibits
distinct behaviors depending on $\xi $. Typical behaviors are as follows.
When $\xi =0$, the localized state rapidly spreads as in Fig.\ref%
{FigPropagate}(a). On the other hand, when $\xi =4$, we observe a
soliton-like wave propagation as in Fig.\ref{FigPropagate}(b). When $\xi =10$%
,\ the state remains localized as in Fig.\ref{FigPropagate}(c). We explore
these characteristic phenomena more in detail.

\textit{Linear model}: The time evolution of the LDOS $\left\vert \psi
_{n}\right\vert ^{2}$ for the linear model ($\xi =0$) is shown in Fig.\ref%
{FigQuench}(a1)$\sim $(a5). The amplitude spreads not only along the edge
but also into the bulk in a symmetric way between the right and the left
sides. One might expect a one-way propagation as in the Haldane model
because the edge states are chiral. However, this is not the case in the
present model, because there are two pairs of chiral edge states with
opposite chiralities as shown in Fig.\ref{FigRibbon}(a). In fact, the
occupation of these two opposite chiral edge states is identical as shown in
Fig.\ref{FigSquare}(b). Namely, there are equal numbers of the right-going
and left-going propagating waves, which results in the symmetric spreading
of the wave propagation. Another feature is that considerable amounts of the
amplitude $\left\vert \psi _{n}\right\vert ^{2}$ penetrate into the bulk
although we start with the state localized at the edge as in Eq.(\ref{IniCon}%
). This is due to the fact that $\left\vert c_{\alpha }\right\vert ^{2}$ is
nonzero also for bulk states as we have already pointed out: See Fig.\ref%
{FigSquare}(b).

\textit{Chiral edge soliton}: We next study the nonlinear model with $\xi =4$%
, whose quench dynamics is shown in Fig.\ref{FigChiral}(a1)$\sim $(a8) for a
square with size $20\times 20$ and in Fig.\ref{FigChiral}(b1)$\sim $(b8) for
a rectangle with size $80\times 10$. There are two features absent in the
linear model. One is that the wave packet propagates rightward. Namely, the
wave packet dynamics becomes chiral due to the nonlinear term. A remarkable
feature is that the shape of the wave packet remains almost unchanged after
a certain time. It is the case even after the wave packet turns a corner, as
shown in Fig.\ref{FigChiral}(a4) and (a8). It looks like a solitary wave or
a soliton. Hence, it may well be a chiral edge soliton. We may interpret
that an initial pulse excites a chiral edge soliton and other modes, where
all the other modes spread into the bulk and disappear.

A comment is in order for the presence of slight oscillations in the
propagation of a solitonic wave. This is due to a lattice effect associated
with the lack of the continuous translational symmetry. The form of a
soliton slightly deforms depending on the mismatch between the center of the
wave packet and lattice points\cite{Muk}. This is in contrast to the
continuum theory, where a soliton moves without deformation.

\begin{figure}[t]
\centerline{\includegraphics[width=0.48\textwidth]{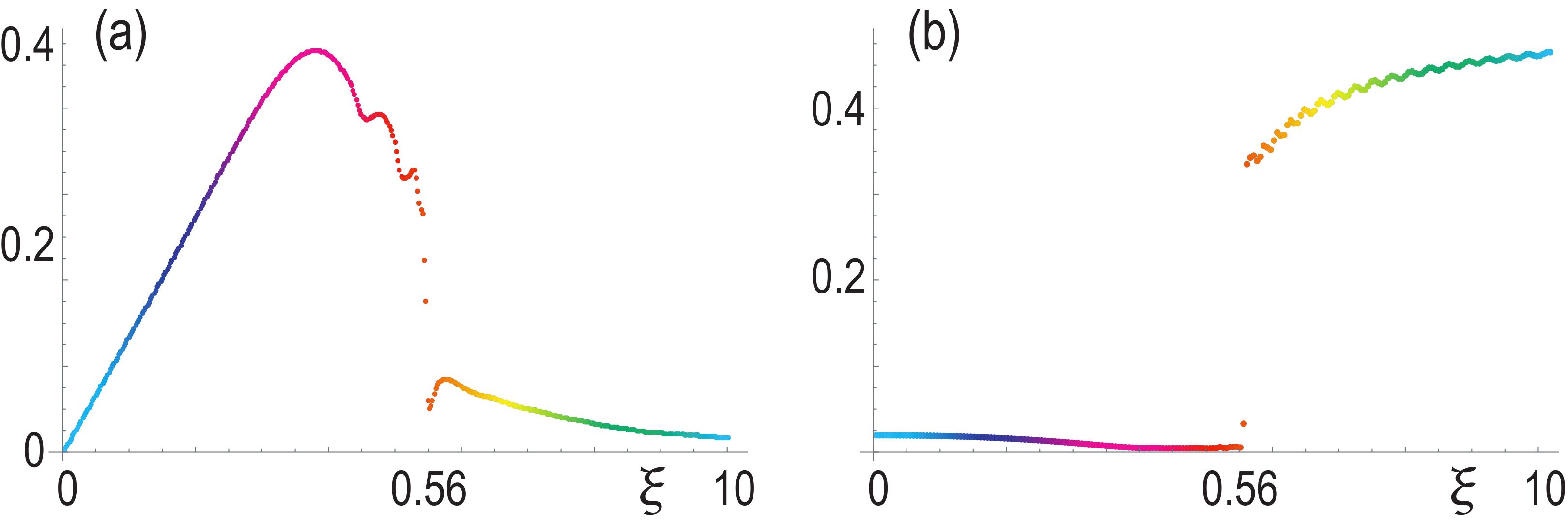}}
\caption{(a) Velocity $v_{x}$ as a function of $\protect\xi $, which is
linear for $\protect\xi \lesssim 4$. The vertical axis is velocity in the
unit $\protect\kappa $, while the horizontal axis is $\protect\xi $. (b)
Amplitude $|\protect\psi _{L_{x}/2,1}|^{2}$ at the initial point as a
function of $\protect\xi $. Color indicates the value of $\protect\xi $. The
horizontal axis is $\protect\xi $. We have used a sample with the size $%
40\times 10$.}
\label{FigFit}
\end{figure}

The mean position $\left\langle x\right\rangle $\ of the wave packet is
given by 
\begin{equation}
\left\langle x\right\rangle \equiv \sum_{n_{x},n_{y}}\left(
n_{x}-L_{x}/2\right) \left\vert \psi _{n_{x},n_{y}}\right\vert ^{2}.
\end{equation}%
We calculate the time evolution of $\left\langle x\right\rangle $ for
various $\xi $, fit the position by a linear function $\left\langle
x\right\rangle =v_{x}t$, and estimate the velocity $v_{x}$ as a function of $%
\xi $, whose result is summarized in Fig.\ref{FigFit}(a). The velocity is
zero for $\xi =0$. It linearly increases for $0<\xi \lesssim 4$, suddenly
decreases for $\xi \gtrsim 4$, and makes a jump at $\xi \approx 5.6$.

\textit{Self-trapped state}: The jump indicates a self-trapping transition.
We show the time evolution of the spatial distribution in a strong nonlinear
model ($\xi =10$) in Fig.\ref{FigQuench}(b1)$\sim $(b5), where the state
remains localized. In order to show the self-trapping transition, we
calculate the amplitude at the initial site after enough time,%
\begin{equation}
\left\vert \psi _{0}\right\vert ^{2}\equiv \lim_{t\rightarrow \infty
}\left\vert \psi _{L_{x}/2,1}\left( t\right) \right\vert ^{2}.
\end{equation}%
We show it as a function of $\xi $ in Fig.\ref{FigFit}(d). There is a sharp
transition around $\xi \simeq 5.6$. The nonlinear-induced self-trapping
transition has been discussed in other contexts\cite{Cai,NLPhoto,NLSkin}.

In the strong nonlinear regime ($\xi \gg 1$), we may approximate Eq.(\ref%
{BasicEq}) as%
\begin{equation}
i\frac{d\psi _{n}}{dt}=-\xi \left\vert \psi _{n}\right\vert ^{2}\psi _{n},
\end{equation}%
where all equations are separated one another. The solution is $\psi
_{n}\left( t\right) =r_{n}e^{i\theta _{n}\left( t\right) }$, with a constant 
$r_{n}$ and $\theta _{n}=\xi r_{n}^{2}t+c$. Hence, the amplitude does not
decrease. By imposing the initial condition (\ref{IniCon}), we have $r_{n}=1$
for $t=0$. Namely, the state is strictly localized at the initial site as in
Fig.\ref{FigPropagate}(c).

\textbf{Discussion}: We have studied nonlinear effects on the chiral edge
state in the nonlinear Schr\"{o}dinger equation. We have revealed
characteristic features depending on the magnitude of the nonlinearity. When
the nonlinearity is negligible, an initial pulse spreads symmetrically and
diffuses. When it is intermediate, a chiral edge soliton is formed from an
initial pulse. When it is strong enough, the self-trapping state appears.

The nonlinear term is an interaction term, which is diagonal in the real
space. On the other hand, the hopping matrix $M$ is diagonal in the momentum
space as in Eq.(\ref{matrixM}). Hence, there is a competition between the
real-space and momentum-space diagonalizations in the nonlinear Schr\"{o}%
dinger equation. In the linear model, the momentum is a good number, where
the plane wave is an eigenstate. On the other hand, in the strong nonlinear
model, the real space is a good number, where the self-trapping state is an
eigenstate. A chiral soliton emerges between these two limits.

It will be possible to observe these phenomena experimentally in various
nonlinear topological systems. The photonic system is a good candidate,
where the nonlinear term is generated as the Kerr term, and the topological
edge states must be directly observed by photoluminescence. It is also
possible to observe the time evolution of the edge states\cite%
{Hafezi,Bandres}. Furthermore, the nonlinearity is controlled only by
changing the intensity of light without changing a sample.

The author is very much grateful to N. Nagaosa for helpful discussions on
the subject. This work is supported by the Grants-in-Aid for Scientific
Research from MEXT KAKENHI (Grants No. JP17K05490 and No. JP18H03676). This
work is also supported by CREST, JST (JPMJCR16F1 and JPMJCR20T2).

\end{document}